\documentclass[%
reprint,
groupedaddress,
 amsmath,amssymb,
 aps,
]{revtex4-1}

\usepackage{graphicx}

\usepackage{subcaption}

\def\D{\mathrm{d}}

\date{\today} 

\begin{document}

\title{Dissipation of 
traffic congestion using agent-based car-following model with modified optimal velocity} 


\author{Manit Klawtanong}
\affiliation{Department of Physics, Faculty of Science, Ramkhamhaeng University, Ramkhamhaeng Road, Bang Kapi, Bangkok 10240}
\author{Surachate Limkumnerd}
\email[]{surachate.l@chula.ac.th}
\affiliation{Physics of Energy Materials Research Unit, Department of Physics, Faculty of Science, Chulalongkorn University, Phayathai Road, Patumwan, Bangkok 10330, Thailand}
\affiliation{Thailand Center of Excellence in Physics, CHE, 328 Si Ayutthaya Road, Bangkok 10400, Thailand}

\begin{abstract}
We investigate dynamical properties of traffic flow using 
the stochastic car-following 
model with 
modified optimal velocity on circular road. The safety distance following the two-second rule and autonomous vehicles, acting as agents, obeying simple requirements 
are incorporated into the model. 
The dynamic safety distance increases in a light traffic condition where the average driving 
velocity is high, 
while decreases in a dense 
traffic condition in anticipation of slower traffic motion. The results show that 
the presence of the agents can 
enhance overall velocity and traffic current of 
the system, and postpone the 
traffic congestion. In a 
particular phase region, imposing a speed limit enables the system 
to leave the congested flow phase. 
The density-dependent speed limit in agent-free condition is obtained to achieve the optimal traffic flow.
\end{abstract}

\maketitle

\section{\label{sec:intro}Introduction}
The collective 
motion of many-particle systems continues to be one of the most prominent topics 
studied in 
non-equilibrium 
processes~\cite{Helbing2001,Nagatani2002,Chowdhury2005}. 
Asymmetric interactions 
between particles 
are studied in 
transport systems because of their rich phenomena, such as phase transition, self-organization, and scaling behaviour~\cite{Biham1992,Tajima2001}. Generally, 
mathematical modeling with regard 
to transport systems can be classified 
into one of the following three categories. i) Microscopic models are 
used to describe collective motion of 
individual particles that follow 
predefined transport rules. The cellular 
automata and car-following models 
are examples of this class of models~\cite{Newell1961,Nagel1992}. 
Others serve as pedagogical models that 
can be solved exactly for some 
processes, e.g., asymmetric simple exclusion 
process with open 
boundaries~\cite{Derrida1992}. ii) Mesoscopic models, also known as gas-kinetic models, are based on the 
Maxwell-Boltzmann transport theory 
of gases~\cite{Prigogine1960}. Lastly iii) macroscopic models where cars are coarse-grained and perceived as  continuous fluid flow. 
Shock wave arising from 
discontinuity of 
the compressible fluid is viewed as 
traffic jam travelling backward with respect to the main flow~\cite{Lighthill1955,Flynn2009}. 
Micro-macroscopic links can be found following simple theoretical assumptions~\cite{Lee2001}. 
For many decades, the transport models have been successfully
applied to many systems in 
seemingly different disciplines 
including ant movement~\cite{John2009}, pedestrian flow~\cite{Helbing2000,Tajima2001}, and motions of molecular motors~\cite{Klumpp2003}. 
 
In the area of vehicular transport, 
traffic congestion is one of 
the most pertinent issues that 
has been widely studied. Traffic signal optimization~\cite{Arita2017}, density and delay-feedback 
control~\cite{Woelki2013,Konishi2000}, 
real-time information provision~\cite{Yokoya2004}, and autonomous 
vehicles~\cite{Stern2018} are 
examples of proposed methods to solve the problem. In this work, we use the stochastic 
car-following model with continuous 
spatial-temporal evolution to investigate the system on a circular road. According to the previous work, the safety distance in the optimal velocity function is independent of 
time and is the same for all drivers~\cite{Nagatani2002,Konishi2000,Bando1995}. However, the predefined safety distance is not practical for the drivers because it should depend on the average velocity of the car in the front. 
Normally, the advised safety distance for a driver follows what is known 
as the ``two-second rule". Our goal is to 
propose dynamic safety distance 
based on the two-second rule and
autonomous driving agents 
to dissipate the 
traffic jam. The dynamic distance 
accounts for more realistic individual's response that is not the 
same for every driver, is not 
fixed during the motion, and also depends on 
the driving velocity. As autonomous 
vehicles become gradually more 
available, they will increasingly play more 
central a role in reducing 
the traffic issues. The proposed agents, accounting for autonomous control units, require only basic information that is accessible with the current technology. We will show 
that the decrease in the traffic congestion can be 
achieved by increasing the number of such agents.

Our work is organized as follows. In 
Sec.~\ref{sec:model}, the car-following 
model based on stochastic differential equations is 
presented. Modified optimal 
velocity, 
agents, model parameters, and computer simulation 
methods are described. Simulation 
results consisting of spatial-temporal 
traffic profiles, the average velocity 
as well as the traffic current, 
and the traffic phase diagram in 
steady states are shown in 
Sec.~\ref{sec:result}. Finally, 
we discuss and conclude our 
work in Sec.~\ref{sec:dis}.

\section{\label{sec:model}Car-Following Model}
To simulate the dynamics of traffic phenomena, we adopt the stochastic car-following model 
with optimal velocity (OV)~\cite{Bando1995PRE,Bando1995,Tadaki1998},
\begin{eqnarray}
\label{eq:1}
\D{X}_{i} &=& V_{i}\,\D{T},\\
\D{V}_{i} &=& \frac{1}{\tau}\left[V_{\text{opt}}
\left(\Delta X_{i} \right)\D{T}-
V_{i}\,\D{T}+\sqrt{2D}\,\D{W}_{i}\right],
\label{eq:2}
\end{eqnarray}
where $X_{i}$ and $V_{i}$ are the position and velocity of car $i$, $\tau$ is the driver's 
response time, $\Delta X_{i} = X_{i-1} - X_{i}$ is a headway distance of car $i$ measured from 
the center of car $i$ to the center of the preceding car $i-1$, $V_{\text{opt}}(\Delta X_{i})$ is an optimal velocity depending completely on the headway distance, $D$ is the 
diffusion constant relating to the strength of velocity variations, and the noise term $\D{W}_{i}\propto\sqrt{\D{T}} $ is a Wiener process. Each driver tries to adjust his or her velocity to reach the optimal velocity according to the headway distance. The change in the velocity is compensated by the response time and noise term characterizing individual's behavior. Macroscopically, factor 
$\sqrt{D/\tau}$ is regarded as the velocity deviation or the velocity at which small perturbations of traffic flow travel~\cite{Flynn2009}. Let us introduce two system's parameters: i) car's length $X_{\text{car}}$ and ii) $V_{0}=X_{\text{car}}/\tau$, and define dimensionless parameters:
$t = T/\tau$, $x_{i}=X_{i}/X_{\text{car}}$, $v_{i}=V_{i}/V_{0}$, $\sigma_{0}=(\sqrt{2D/\tau})/V_{0}$, and $\D{w}_{i}=\D{W}_{i}/\sqrt{\tau}$. Equations~(\ref{eq:1}) and (\ref{eq:2}) become
\begin{eqnarray}
\label{eq:drif_diff1}
\D{x}_{i} &=& v_{i}\D{t},\\
\D{v}_{i} &=& \left[v_{\text{opt}}
\left(\Delta x_{i} \right)-
v_{i}\right]\D{t}+\sigma_{0}\D{w}_{i}.
\label{eq:drif_diff2}
\end{eqnarray}
Eq.~(\ref{eq:drif_diff2}) is also known as It{\^o} drift-diffusion process, where the first term on the right hand side is the drift term and the 
second term is the diffusion term. The optimal velocity is an increasing function of $\Delta x_{i}$ with conditions: 
(1) $v_{\text{opt}}
\left(\Delta x_{i} \right) = u_{0}$ 
(the maximum velocity) when $\Delta x_{i}\rightarrow\infty$ and (2) $v_{\text{opt}}\left(\Delta x_{i} \right) = 0$ when $\Delta x_{i}\leq\Delta x_{\text{min}}$ (some minimum distance). The first condition is to ensure the maximum velocity 
on an empty road, while the second condition is to avoid collision with other cars. A sigmoid 
function~\cite{Bando1995PRE,Tadaki1998} is normally employed in this case:
\begin{widetext}
\begin{equation}
v_{\text{opt}}\left(\Delta x_{i}\right)
=u_{0}\left\{\frac{\tanh\left[
a\left(\Delta x_{i} - \Delta x_{\text{safe}}-\Delta x_{\text{min}}\right)\right]+
\tanh\left[a\left(\Delta x_{\text{safe}}\right)\right]}
{1+\tanh\left[a\left(\Delta x_{\text{safe}}\right)\right]}\right\}
\label{eq:v_opt0}
\end{equation}
\end{widetext}
where safety distance $\Delta x_{\text{safe}}$ and $a$ are model parameters. Minimum distance $\Delta x_{\text{min}}$ is set to be $1$ (corresponding to the distance of one car), thus $\Delta x_{i} -  
\Delta x_{\text{min}}$ equals a distance 
from the front bumper 
of car $i$ to the rear bumper of car $i-1$.

\subsection{Modified optimal velocity}
In this work, safety distance $\Delta x_{\text{safe}}$ is replaced by a 
method of ensuring a safe headway, 
commonly known as the ``two-second rule" 
\begin{equation}
\Delta x_{\text{safe},i} = \bar{v}_{i-1}\Delta t_{\text{c}},
\label{eq:dx_safe}
\end{equation}
where $\Delta 
t_{\text{c}}=2~\text{seconds}/\tau$ is the recommended time for a driver to 
stay behind the car in the front to ensure that he or she receives enough time to respond and 
$\bar{v}_{i-1}$ is the temporal average velocity that the driver of car $i$ perceives of the preceding car. Note that we model the system on a circular road, 
so $\bar{v}_{N+1} = \bar{v}_{1}$. The remaining parameter 
to be specified is $a$. The sensitivity of the 
OV function with respect to small change $\Delta x_{i}$, i.e. 
$\D{v}_{\text{opt}}/\D{\Delta x_{i}}$, 
has the maximum value at $\Delta x_{i} = \Delta x_{\text{safe},i} +\Delta x_{\text{min}}$. We set the shape, or more 
precisely the full width at half maximum (FWHM), of the 
sensitivity to be scaled with its corresponding safety distance $\Delta x_{\text{safe},i}$. This means that 
optimal velocity ${v}_{\text{opt}}$ becomes less sensitive to 
headway distance $\Delta x_{i}$ when the safety distance increases. Parameter $a$ can then be determined from $\text{FWHM}=2\cosh^{-1}\left(\sqrt{2}\right)/a=\alpha \Delta x_{\text{safe},i}$, where $\alpha$ is a dimensionless scaling constant.   

\subsection{Agents}
In this section, we introduce an autonomous car or an agent and implement a feasible method for agents to damp traffic congestion. An agent is a self-driving 
car equipped with a control unit. The Cognitive and Autonomous Test (CAT) vehicle~\cite{Stern2018} is one 
example of such an agent. 
To differentiate agent with (normal) car driving behavior, a few simple rules are required for the agents. First, when a car in the front changes its 
velocity, the agent can react instantaneously and 
reaches its desired optimal velocity without any delay~\footnote{We note that, in real driving experience, a short response time is required 
for safety and comfortable driving.}. 
Second, there are no fluctuations in the changing velocity, i.e. the noise term in Eq.~(\ref{eq:drif_diff2}) is zero. 
Small fluctuations resulting from human driving behavior 
are known to cause `stop-and-go' wave to occur 
even from initially homogeneous 
motion~\cite{Sugiyama2008,Stern2018}. Finally, since 
autonomous driving is safer than manual driving, the 
agent can cruise at a shorter safety distance. Here we set $\Delta t_{\text{a}}$ to half that of the normal vehicles. For simplicity, the agents do not communicate with 
each other and have the same information 
as the normal drivers. 
There is no information, for example, about an incoming 
traffic wave, sharing among the agents.

\subsection{Parameters and simulations}
Here, we describe model parameters and simulation methods. We investigate the car-following model on a periodic road with the total distance of $l=100$ (equal to 100 cars). The 
normal car and agent densities are given by 
$\rho_{\text{c}}=n_{\text{c}}/l$ and $\rho_{\text{a}}=n_{\text{a}}/l$, where $n_{\text{c}}$ and $n_{\text{a}}$ are the numbers of normal cars and agents, respectively. The total number of vehicles on the road is $n=n_{\text{c}}+n_{\text{a}}\leq l$ 
corresponding to 
the total vehicle density of $\rho_{\text{t}}=n/l=\rho_{\text{c}}+\rho_{\text{a}} \leq 1$. Agents are placed randomly on the circular track. Imposed 
condition $\Delta x_{i}$ (or $\Delta x_{\text{safe},i}$) $=\Delta x_{\text{min}}$ is used 
whenever $\Delta x_{i}$ (or $\Delta x_{\text{safe},i}$) $<\Delta x_{\text{min}}$ so that 
$v_{\text{opt}}\left(\Delta x_{i}\right)=0$. The 
changes in velocity and position at time $t$ 
in Eqs.~(\ref{eq:drif_diff1}) and (\ref{eq:drif_diff2}) 
are updated with fixed small time step 
$\D{t} = 0.1$, corresponding to the physical time of about $50$~ms. 
At time $t'=t+\D{t}$, velocity $v_{i}(t') = 
v_{i}(t) + \D{v}_{i}(t)$ is modified for each vehicle within the range, $0 \leq v_{i}(t') \leq u_{0}$. Then spatial position $x_{i}(t') = x_{i}(t) + \D{x}_{i}(t)$ is updated sequentially such that each 
car moves on the road without 
interfering with the car in the front.

The position and velocity of each car are tracked and 
measured. The average velocity at time $t$ is given by 
\begin{equation}
v_{\text{av}}(t)=\frac{1}{n}\sum_{i=1}^{n}v_{i}(t).
\label{eq:vel}
\end{equation}
We use the standard deviation $\sigma_{v}(t)$ of all car velocities  
 to determine whether or not 
small initial perturbation $\sigma_{0}$ will be amplified 
during the motion. If $\sigma_{v}$ is greater than 
predetermined value, 
$\sigma_{\text{max}} = c\sigma_{0}$, the 
motion is considered to be congested. Constant $c=\sqrt{2}$ corresponds to 
$\sigma_{\text{max}}V_{0}=3.0~$m/s, which is close to the threshold of $2.5~$m/s used in the field experiments by Stern et al.~\cite{Stern2018} to identify the onset of traffic wave. If congestion occurs, jamming state 
$J$ is said to be $1$, otherwise 
it is zero. 
Both $v_{\text{av}}$ and $J$ are averaged over the course of 1,000 and 500 trials, respectively. 
Finally if more than $50\%$ of the trials belong to state $J=1$, we justify the 
traffic condition to be in the congestion region. Some 
model constants are listed in Table~\ref{table:param}. 
\begin{table}[t]
\caption{Constants used in this work.}
\centering
 \begin{tabular}{|c|c|c|c|} 
 \hline
 parameter & value & units & Refs.\\  
 \hline\hline
 $\tau$ & 0.5 & second (s) & \cite{Tadaki1998,Bando1995}\\ 
 $X_{\text{car}}$ & 5.0 & meter (m) & \cite{Bando1995}\\
 $\sqrt{D/\tau}$ & 1.5 & m/s &\\
 $\Delta x_{\text{min}}$ & 1.0 & - &\\
  $\alpha$ & 0.5 & - &\\
   $\Delta t_{\text{c}}$ & 4.0 & - &\\
   $\Delta t_{\text{a}}$ & 2.0 & - &\\ 
 \hline
 \end{tabular}
 \label{table:param}
\end{table}
 
\section{\label{sec:result}Simulation Results}

\subsection{Traffic profiles}
\begin{figure*}[t]
    \centering
	    \begin{subfigure}[t]{0.32\textwidth}
	        \centering
	        \includegraphics[trim={0 0mm 8mm 0},clip=true,scale=0.41]{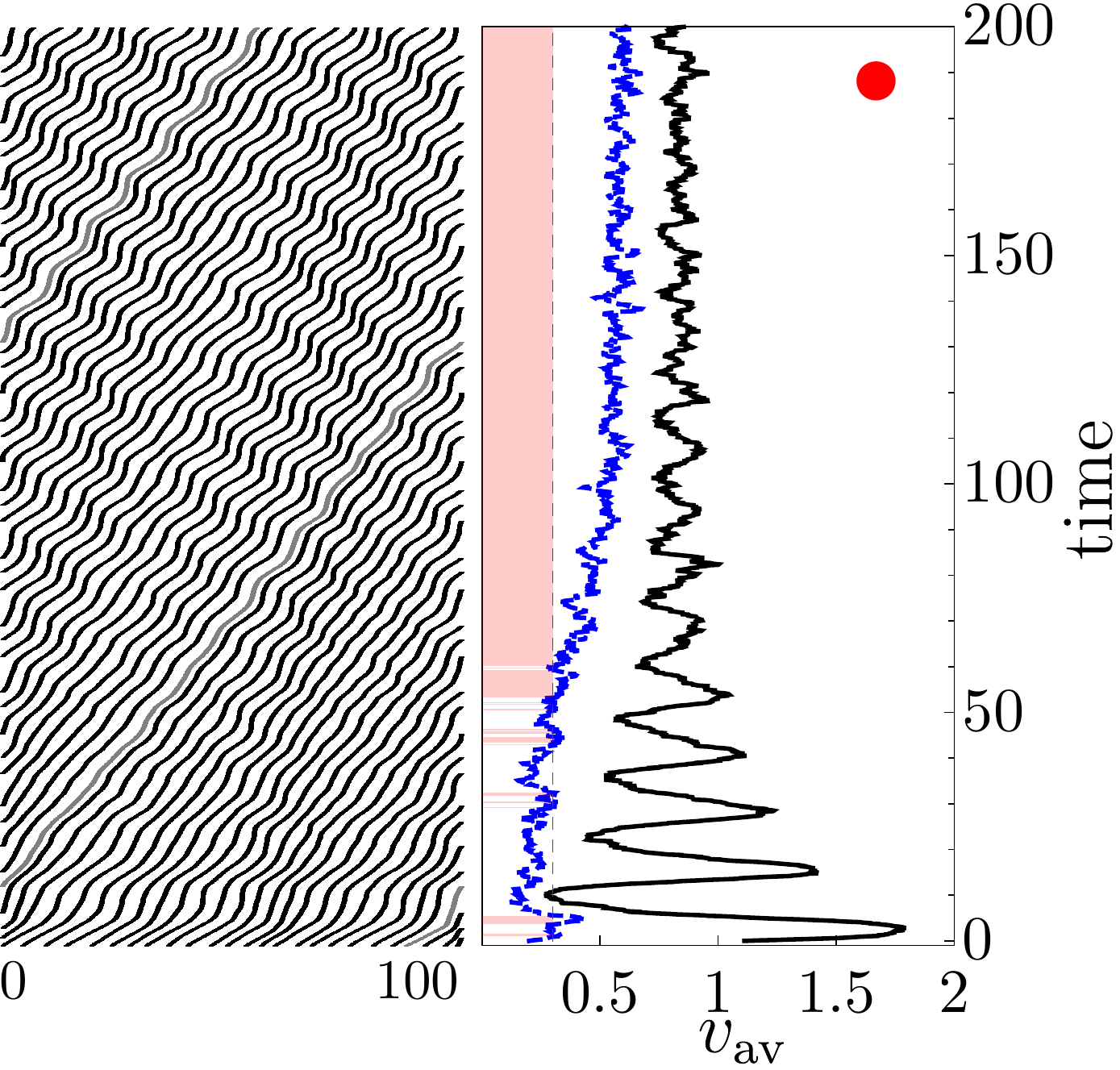}
	        \caption{$\rho_{\text{a}}=0.01$.}
	    \end{subfigure}%
	    ~ 
	    \begin{subfigure}[t]{0.32\textwidth}
	        \centering
	        \includegraphics[trim={0 0mm 8mm 0},clip=true,scale=0.41]{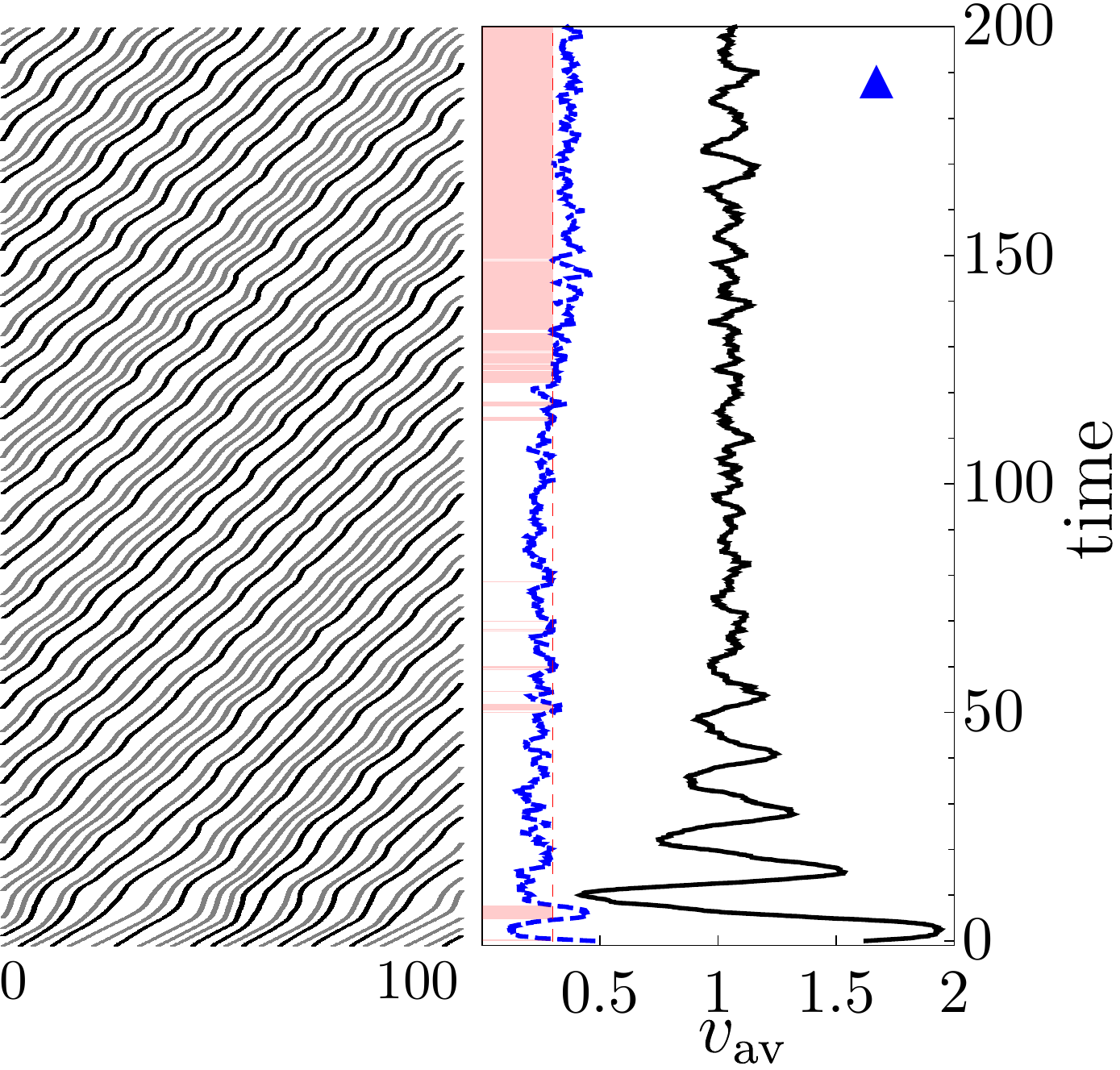}
	        \caption{$\rho_{\text{a}}=0.15$.}
	    \end{subfigure} 
	    ~ 
	    \begin{subfigure}[t]{0.32\textwidth}
	        \centering
	        \includegraphics[scale=0.41]{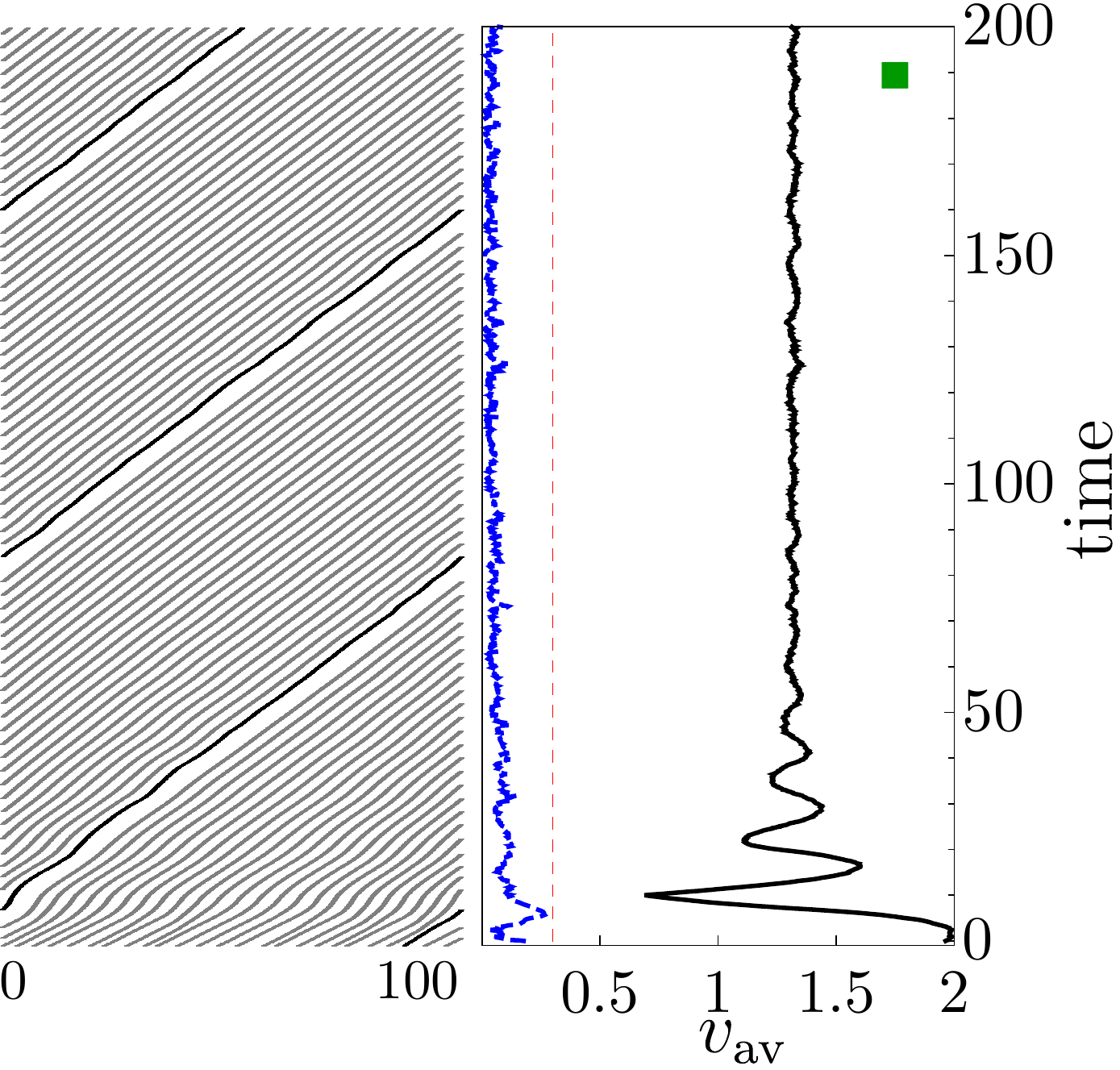}
	        \caption{$\rho_{\text{a}}=0.24$.}
	    \end{subfigure}
    \caption{Trajectory profiles at total density $\rho_{\text{t}}=0.25$ with different agent 
    densities: (a) $\rho_{\text{a}} = 0.01$, 
    (b) $\rho_{\text{a}} = 0.15$, and 
    (c) $\rho_{\text{a}} = 0.24$. The normal car (agent) trajectory, moving from the left to the right, is indicated by the black (gray) line. Time direction 
    increases from the bottom to the top. Corresponding average velocity $v_{\text{av}}$ (thick solid line) and 
    standard deviation $\sigma_{v}$ (thick dashed line) are shown on the 
    right hand side of each profile. Areas under 
    the congestion 
    threshold $\sigma_{\text{max}}$ 
    (thin dashed line) are highlighted if 
    $\sigma_{v}>\sigma_{\text{max}}$. The filled circle, 
    up-triangle, and square symbols are used to 
    mark traffic 
    conditions of the profiles in different phase regions.}
    \label{fig:profiles}
\end{figure*}
We first show simulation results with maximum 
velocity of $u_{0} = 2.0$ (or about $72$~km/hr), unless otherwise 
stated. Fig.~\ref{fig:profiles} shows 
three space-time profiles of car trajectories. 
The horizontal axis indicates car positions moving 
from left to right. The vertical axis shows the direction of time increasing from bottom to top. 
Each black (gray) line represents a car's (an agent's) 
trajectory. On the right hand side of each profile, 
average velocity $v_{\text{av}}$ (solid line) and 
standard deviation $\sigma_{v}$ (dashed line) 
are shown. The thin dashed line indicates congestion 
threshold $\sigma_{\text{max}}$. 
Highlighted areas denote when congestion occurs 
($\sigma_{v}>\sigma_{\text{max}}$).  The filled circle, 
up triangle, and square, shown at the upper 
right corner, represent 
particular conditions corresponding to the profiles 
that illustrate traffic motions in different phase 
regions which will be described in detail later in the next section. The total number of vehicles is kept at 
$\rho_{t}=0.25$. At the initial time, the car velocity is set closely 
to the optimal velocity. 

The transient oscillations of $v_{\text{av}}$ 
are clearly seen before the traffic flow reaches 
non-equilibrium steady states. For 
$\rho_{\text{a}}\leq 0.15$ 
(Figs.~\ref{fig:profiles}(a) and (b)), 
initial noises are enhanced such that $\sigma_{v}$ 
exceeds threshold $\sigma_{\text{max}}$. At the steady state, the onset of permanent traffic congestion occurs. The characteristic shock wave traveling backward is clearly visible. 
For $\rho_{\text{a}} = 0.24$ 
(Fig.~\ref{fig:profiles}(c)), on the other hand, 
initial perturbations are suppressed. Average velocity $v_{\text{av}}$ increases substantially 
at the steady state. For a small (large) number of agents, the traffic flow belongs to the jamming $J=1$ (free $J=0$) state. At a moderate agent density, 
e.g. $\rho_{\text{a}}=0.15$ the flow lies close to 
the boundary region with state $J$ being either $0$ or $1$. Compared to the results where the agents are absent, 
$v_{\text{av}}$ increases by 2\%, 26\%, 
and 57\% for $\rho_{\text{a}}=0.01$, $0.15$, 
and $0.24$, respectively. It is noted that for $\rho_{\text{a}}=0.24$ even a few drivers may reduce the traffic flow overall.

\begin{figure*}[t]
    \centering
    \begin{subfigure}[t]{0.32\textwidth}
        \centering
        \includegraphics[scale=0.47]{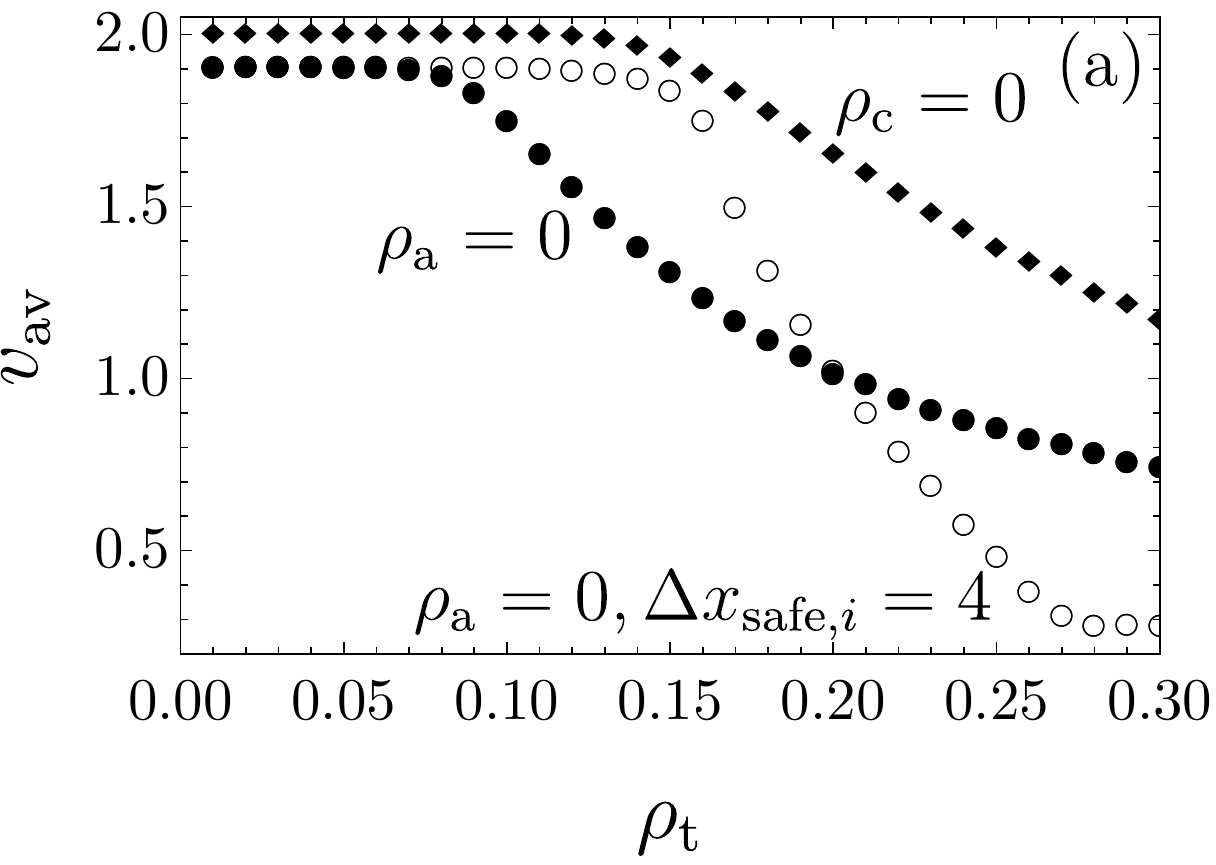}
    \end{subfigure}%
~ 
    \begin{subfigure}[t]{0.32\textwidth}
        \centering
        \includegraphics[scale=0.47]{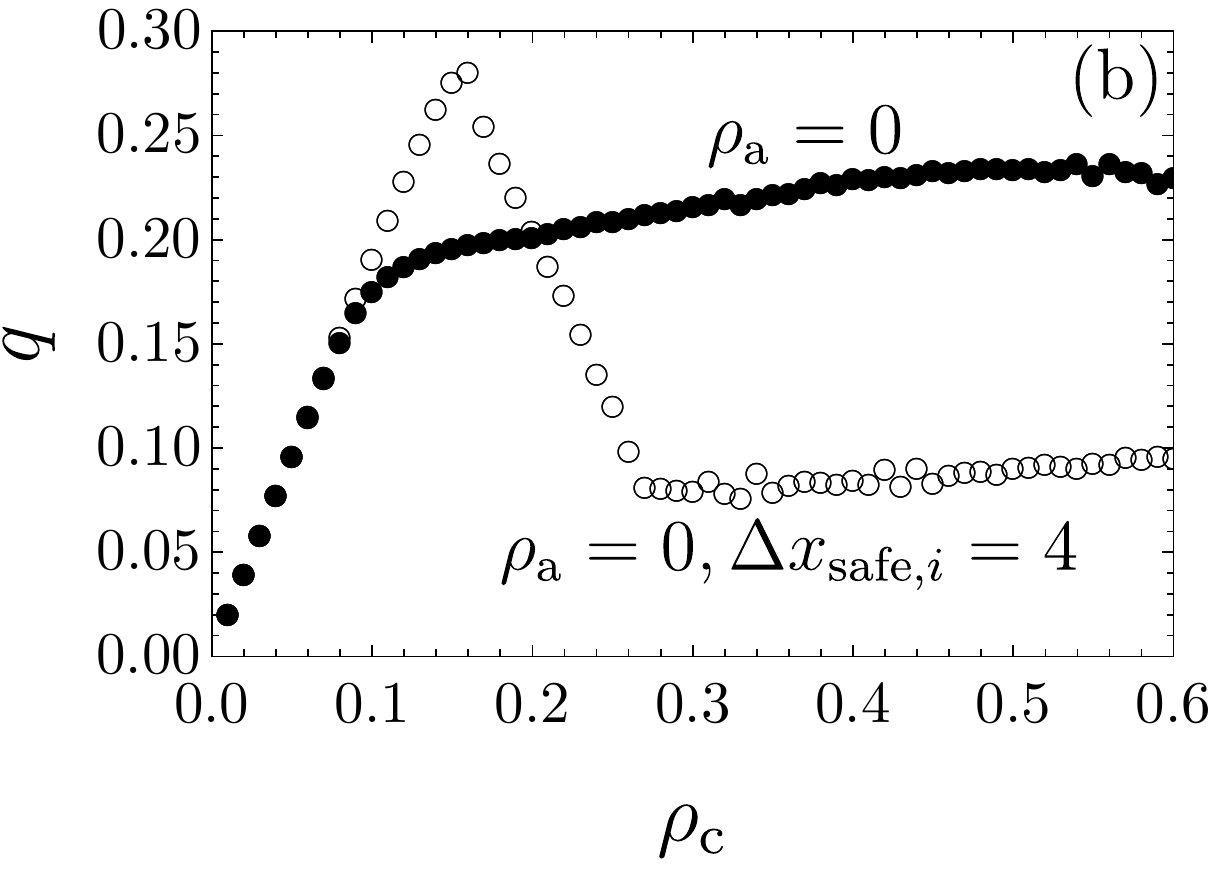}
    \end{subfigure} 
    ~ 
    \begin{subfigure}[t]{0.32\textwidth}
        \centering
        \includegraphics[scale=0.47]{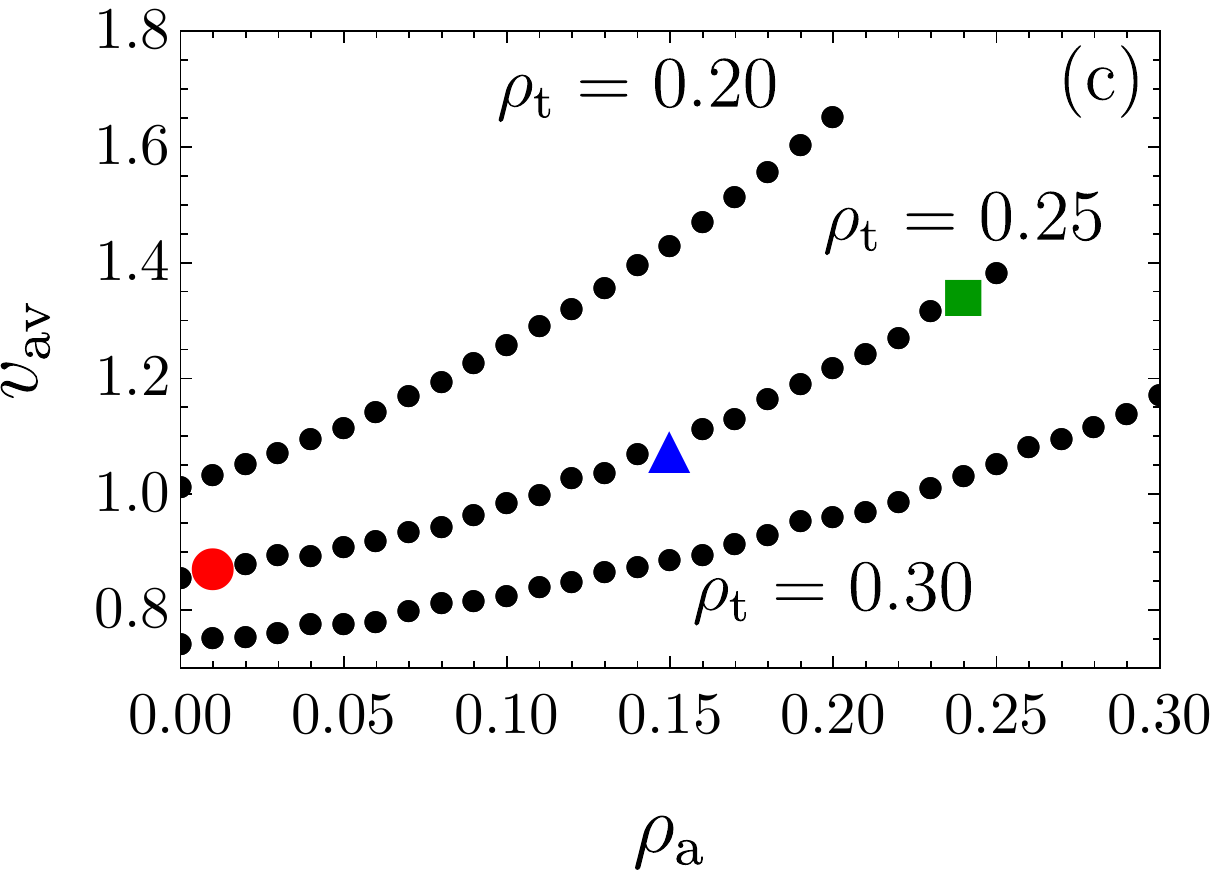}
    \end{subfigure} 
    \caption{(a) Average velocity $v_{\text{av}}$ versus total density $\rho_{\text{t}}$ for $\rho_{\text{a}} = 0$ 
    (empty and filled circles)
    and $\rho_{\text{c}}=0$ 
    (filled diamonds). The empty circle data 
    corresponds to OV with fixed 
    $\Delta x_{\text{safe},i} = 20~\text{meters}/X_{\text{car}}=4$. (b) The traffic current, $q = \rho v_{\text{av}}$, 
    versus car density $\rho_{\text{c}}$ of 
    those results with $\rho_{\text{a}} = 0$ 
    in Fig.~\ref{fig:average_vel}(a). (c) Average velocity $v_{\text{av}}$ versus agent 
    density $\rho_{\text{a}}$ 
    for $\rho_{\text{t}} = 0.20$,
     $0.25$, and $0.30$. 
    The filled circle, 
    up-triangle, and square symbols at 
    $\rho_{\text{t}} = 0.25$ correspond to the profiles in Fig.~\ref{fig:profiles}.}
\label{fig:average_vel}
\end{figure*}

To study the traffic flow at the steady 
states, velocities of individual vehicles are 
averaged after simulation time $t=50$ 
for various values of traffic density. 
Fig.~\ref{fig:average_vel}(a) 
shows $v_{\text{av}}$ versus $\rho_{\text{t}}$ plot of the limiting cases 
where only normal drivers are present 
on the road, i.e., $\rho_{\text{a}} = 0$ 
(empty and filled circles) with 
total density $\rho_{\text{t}}$ being 
$\rho_{\text{t}} = \rho_{\text{c}}$ and 
only agents on the road, $\rho_{\text{c}} = 0$ 
(filled diamonds) with $\rho_{\text{t}} = \rho_{\text{a}}$. For the former case with $\rho_{\text{a}} = 0$, filled circles 
are the simulation results with modified OV, 
while empty circles are those with static 
safety distance 
$\Delta x_{\text{safe},i}$ being set to 
4.0 corresponding to physical safety distance of 
20~meters~\cite{Bando1995}. 
With modified OV, the average velocity 
is maintained at $v_{\text{av}}\simeq 
1.9$ up to  $\rho_{\text{t}}$ of about $8\%$  
before $v_{\text{av}}$ drops 
continuously at a higher density. It is noted that at the low traffic region, 
each driver does not 
always cruise at the maximum velocity of 
$u_{0}=2.0$ because of small 
fluctuations introduced in the system, thus $v_{\text{av}} < u_{0}$. Although 
$v_{\text{av}}$ decreases sooner 
than that with fixed 
$\Delta x_{\text{safe},i}$, the rate at which 
it decreases is lower. 

Average velocity $v_{\text{av}}$ reaches the maximum velocity $u_{0}$
when there are only agents on the road in 
the low traffic region due to the absence of 
noises in the system. The average velocity then slowly decreases 
at a higher car density to compensate 
for the smaller headway distance. The traffic flow is 
improved for all traffic conditions, 
particularly in the heavy traffic.
Compared with the results without any agents ($\rho_\text{a}=0$), systems with agents show moderate (5\% increase in average speed at $\rho_{\text{t}}=0.01$, for example) to remarkable speed boost (64\% increase at $\rho_{\text{t}}=0.22$). 
The results suggest the 
obvious benefits of the agents in diminishing 
the traffic congestion.  

Fig.~\ref{fig:average_vel}(b) shows 
traffic current $q \equiv \rho v_{\text{av}}$ of 
the results with the absence of agents. For 
OV with static $\Delta x_{\text{safe},i}=4$ 
(empty circle data), the sharp change with a possible 
discontinuity~\cite{Bando1995} connecting free flow and 
congested flow phases is visible. Current 
$q$ in the heavy traffic is not zero 
and is again due to the effect of the diffusion term. Although the modified OV model with the two-second rule (filled circle data) does not improve 
the traffic current much 
at low $\rho_{\text{c}}$; it is even worse at the boundary, 
the model enhances the traffic current considerably  when the traffic becomes congested at high $\rho_{\text{c}}$. Note that 
a sharp change in $q$ is not visible in this result.

Traffic flow can be improved further by 
turning the cars to the agents as illustrated 
in Fig.~\ref{fig:average_vel}(c). The trend 
is similar for all conditions of 
total density $\rho_{\text{t}}$. For 
each line, the result is bounded between its 
minimum at $\rho_{\text{a}}=0$ and 
its maximum 
at $\rho_{\text{c}}=0$ as shown 
in Fig.~\ref{fig:average_vel}(a) for a given $\rho_{\text{t}}$. The 
symbols: filled circle, up-triangle, and 
square represent those conditions whose 
traffic profiles are shown in Fig.~\ref{fig:profiles}.

\subsection{Phase diagram}

\begin{figure}[t!]
\centering
\includegraphics[width=0.47\textwidth]{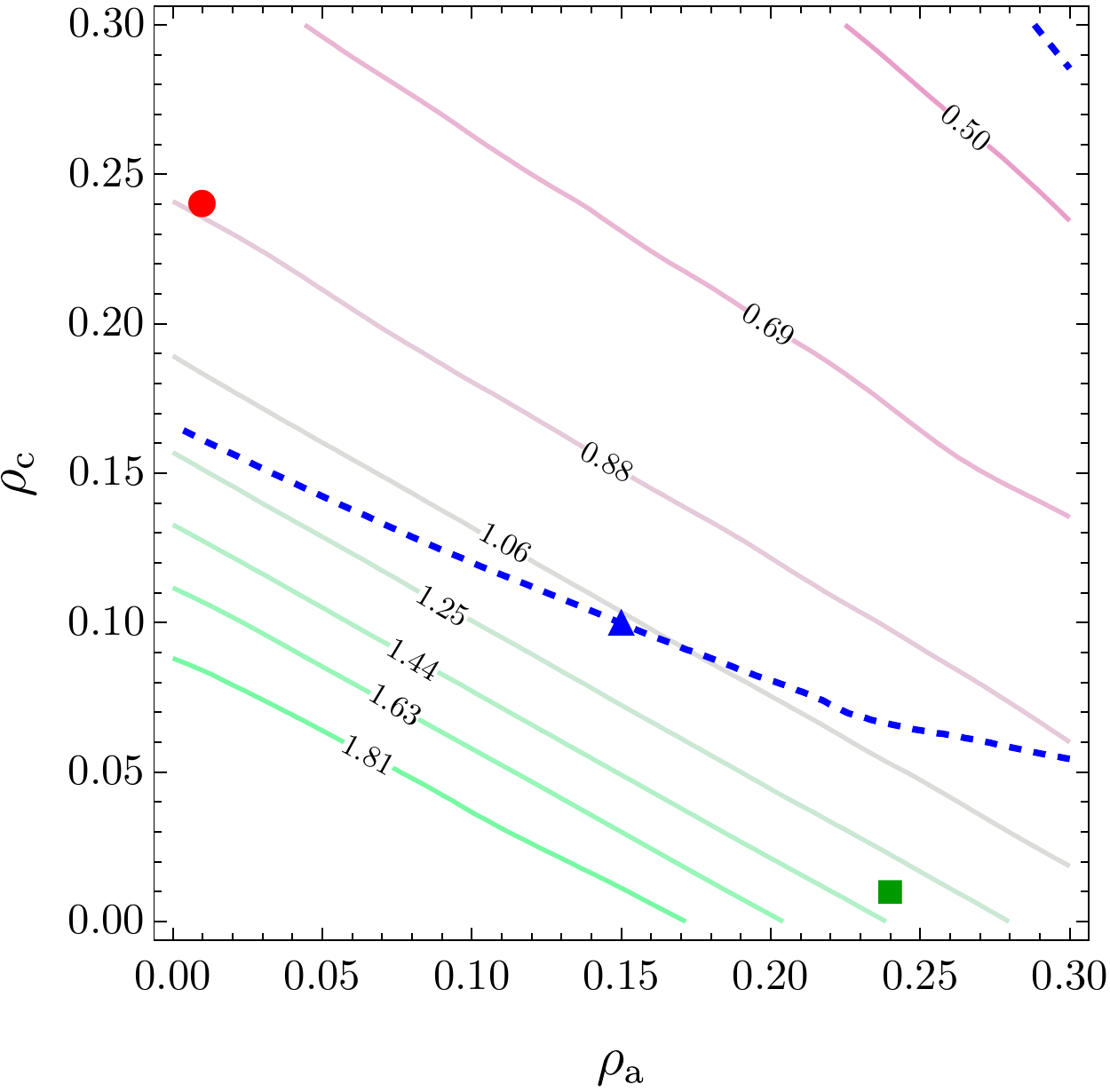}
\caption{
Contour plot of $v_{\text{av}}$ 
as a function of normal car density 
$\rho_{\text{c}}$ and agent density 
$\rho_{\text{a}}$ with limit 
velocity $u_{0}=$ 2.0. 
The dashed 
line indicates a traffic congestion boundary between free flow and congested flow. The filled squares and circles 
are some of those conditions belonging to free flow 
and congested flow, while the filled 
triangle indicates a condition lying close to 
the boundary.}
\label{fig:traffic_phases}
\end{figure}

We turn our attention to a phase diagram between the free flow and congested flow phases. Fig.~\ref{fig:traffic_phases} shows a 
contour plot of average velocity $v_{\text{av}}$ versus normal car 
density $\rho_{\text{c}}$ and 
agent density $\rho_{\text{a}}$. 
As described earlier, 
$v_{\text{av}}$ is 
improved considerably when all cars 
are agents for 
a given total density $\rho_{\text{t}}$. 
The information of $v_{\text{av}}$ 
alone, however, does not 
completely specified the 
traffic situation. Smooth and 
stable motion relating to 
fuel consumption, braking event, 
and car collision are also 
important parameters of interest, 
which are hinted by the traffic phase diagram. 

If more than half of the trials belong to jamming state $J=1$, 
the traffic condition more likely leads to a congestion. We therefore categorize this condition as belonging to the congested 
flow phase, otherwise it is in 
the free flow phase. The boundary 
separating these two phases are indicated 
by the dashed line in 
Fig.~\ref{fig:traffic_phases}. 
It is obvious that the traffic 
phase depends strongly on the 
number of normal cars on the 
tracks. At $\rho_{\text{c}}=0.05$, for example, 
the traffic is in the free flow phase, regardless of the 
number of agents. Moreover, 
for a given total density 
$\rho_{\text{t}}$ (e.g. $\rho_{\text{t}} = 0.25$), 
increasing the number of agents 
will postpone the traffic jam---from congested flow 
(filled circle data) to the 
boundary (filled triangle data), 
and to the free flow (filled square data). It 
should be noted that
at high density, e.g., 
$\rho_{\text{t}}=0.6$, the system belongs to 
the free flow phase. This 
is because short wavelength 
perturbations are suppressed by 
the narrow headway distance 
such that very little velocity variation takes place ($\sigma_{v}$ does not 
exceed the congestion threshold). 
Even though the traffic is not 
congested, the average velocity 
in the region is 
not very high. This suggests that 
$v_{\text{av}}$ and traffic phase 
should be considered together if 
we want to limit the number of normal cars or to increase 
the number of agents before the 
traffic enters the congested flow 
phase.

\begin{figure}[t!]
\centering
\includegraphics[width=0.47\textwidth]{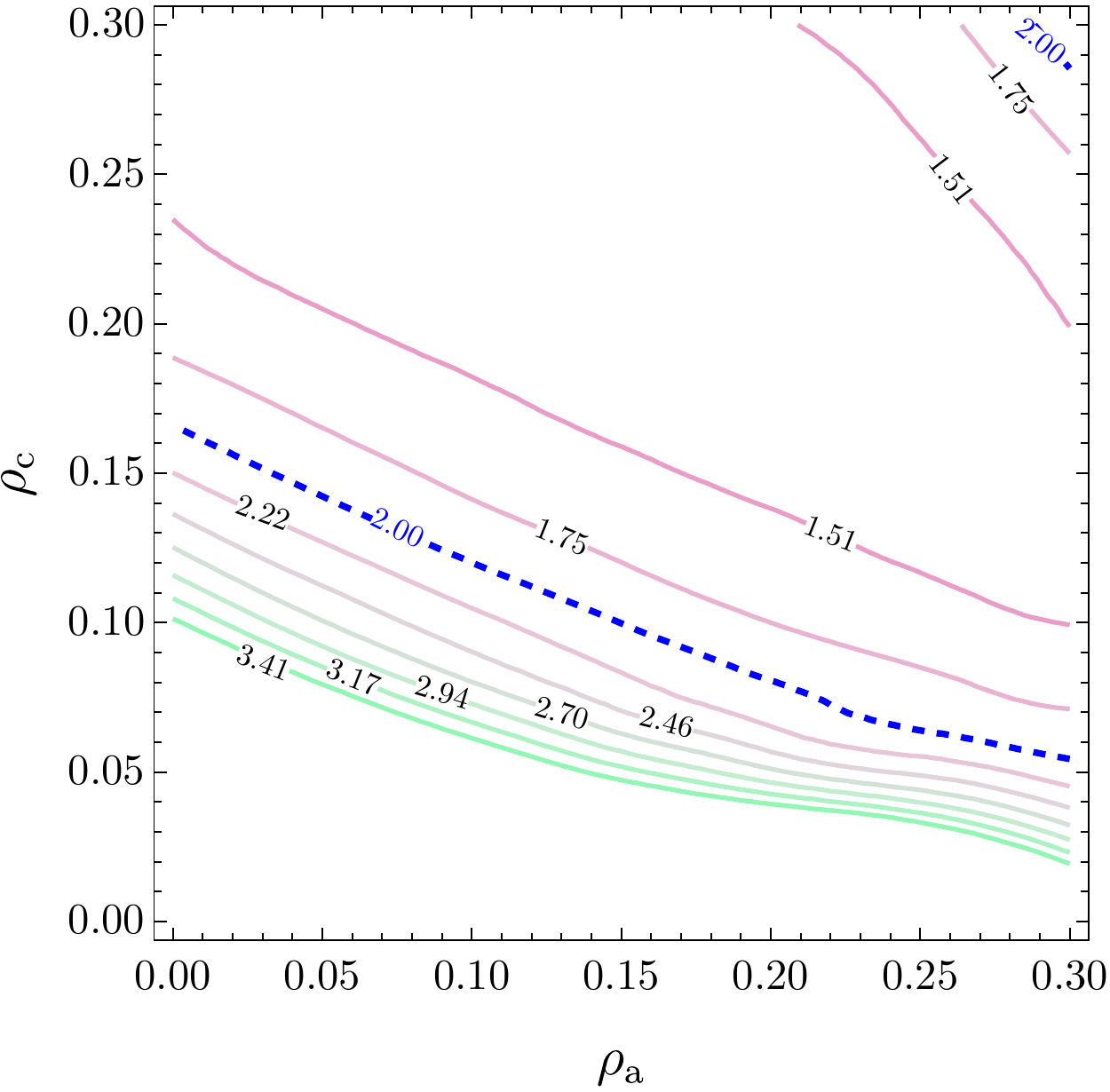}
\caption{Maximum velocity $u_{0}$ 
in a free flow phase as functions 
of normal car density 
$\rho_{\text{c}}$ and agent density 
$\rho_{\text{a}}$. The dashed 
line represents the maximum 
velocity of $u_{0}=2.0$.}
\label{fig:limit_vel}
\end{figure}

In practice, restricting the number 
of cars and/or agents is often not practical. It is however still possible to regulate the free flow by limiting speed limit $u_{0}$. A natural question arises: given 
a traffic condition, what is the speed limit beyond which the traffic becomes congested? 
Fig.~\ref{fig:limit_vel} shows 
the contour plot of the speed limit versus normal car density 
$\rho_{\text{c}}$ and agent 
density $\rho_{\text{a}}$ in the 
free flow phase. The dashed 
line marks the maximum velocity 
of $u_{0}=2.0$. Below the 
dashed line is an area that 
$u_{0}$ can be set higher than 
$u_{0}=2.0$. 
At $\rho_{\text{a}}=0.30$ with 
a few normal cars, for example, 
the maximum velocity can be up to 
$u_{0}=3.41$ corresponding to 
the physical velocity of about 
$120$~km/hr, which exceeds the 
speed limit allowed in many 
countries. Above the dashed 
line is the area where we have to lower 
the speed limit. 
\begin{figure}[t!]
\centering
\includegraphics[width=0.47\textwidth]{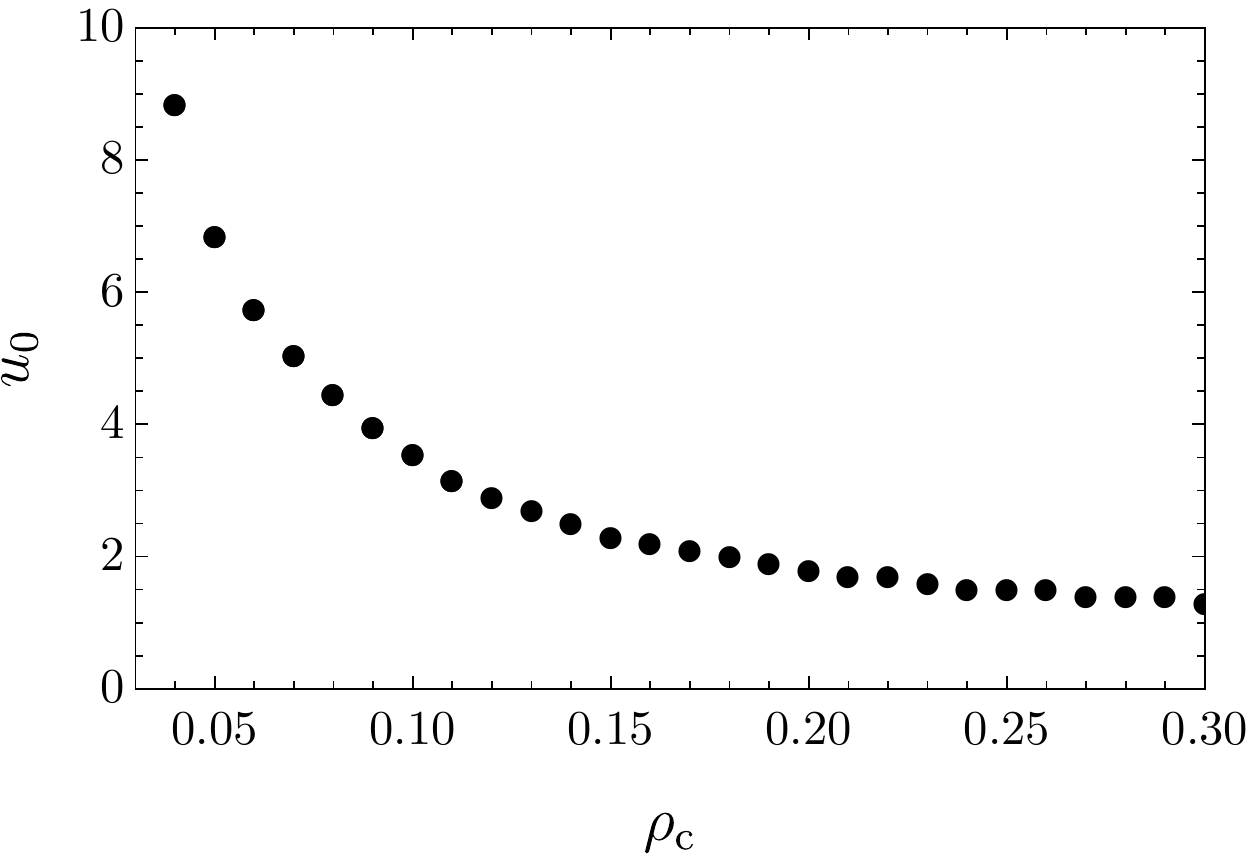}
\caption{Maximum velocity $u_{0}$ 
in a free flow phase with $\rho_{a}=0$ versus normal car density $\rho_{c}$.}
\label{fig:limit_rhoa0}
\end{figure}
Fig.~\ref{fig:limit_rhoa0} shows, 
for instant, variable speed limit 
as a function of car density in 
the absence of agents. As expected, the limit decreases continuously as the car density increases. The result 
can be adapted directly as a new traffic control strategy.

\section{\label{sec:dis}Discussion and Conclusion}
We study the dynamics of the stochastic car-following model with the optimal 
velocity function on a circular road. Two strategies 
are employed in this work. First,
we incorporate the two-second rule 
to determine the safety distance to ensure 
a more realistic driver's response. When we apply the rule 
to the normal drivers (see Fig.~\ref{fig:average_vel}(a) for 
$\rho_{\text{a}}=0$), average 
velocity $v_{\text{av}}$ is 
less than that of the OV with fixed 
$\Delta x_{\text{safe},i}=4.0$ in light traffic condition, while 
it becomes greater at a particular traffic density. The crossover of the 
behavior is due to the fact that 
at low  $\rho_{\text{t}}$, a driver following the two-second 
rule tends to leave a wider safety distance, hence slower 
optimal velocity. At 
a crossover $\rho_{\text{t}}$ where the traffic becomes heavier, 
each driver lowers his or her safety distance 
to compensate for the flow such 
that all cars can continue to move collectively on the crowded road. We intend to modify the optimal velocity in 
this way because when there are several cars on the road, it is generally not safe to drive the car at high speed while keeping the safety distance too small. At dense traffic, 
in contrast, it is not necessary to leave 
a wide safety distance because of the slow, congested traffic. In fact, the cars are nearly bumper-to-bumper. It turns out that the modification of 
the safety distance improves the traffic condition noticeably in a heavy traffic condition.

Second, we 
add to the model autonomous vehicles, acting 
as agents, with a few simple rules. As expected, agents 
increase the average 
velocity of the system for all traffic 
conditions. The presence of the agents not only reduces fluctuations in the system but also 
provides more available road space since the 
agents cruise at a shorter safety distance. Moreover, if the headway distance 
is large enough that it 
looks as if the cars move on the road with low density traffic, uniform motion 
will be stable against 
 small fluctuations. Under such conditions, agents  can definitely dissipate the traffic congestion and increase overall traffic current, particularly 
in the heavy traffic region as 
suggested by the results 
in Fig.~\ref{fig:average_vel}(a). The strategies show promising results, which may additionally 
save more fuel consumption, reduce braking 
events, and car collisions due to the uniform motion of the vehicles~\cite{Stern2018}.

To promote the traffic 
flow with normal cars and 
agents all together, there are two possible 
solutions. 
The first one is to restrict the number of normal cars or increase the number of agents  
to postpone the traffic jam (see Fig.~\ref{fig:traffic_phases}). We note that maximizing 
either average velocity or traffic current alone 
is not adequate, since the traffic may encounter 
the congested state according to the 
results in Fig.~\ref{fig:average_vel}(b). The other one is to adjust 
the maximum velocity for a given 
traffic density such that 
the traffic is still in the free 
flow phase. The imposing limit 
is suggested by the results in Fig.~\ref{fig:limit_vel}. The latter strategy can be implemented immediately in practice, given the ubiquity of traffic cameras from 
which the car density can be calculated.

\begin{acknowledgments}
This research is supported by Rachadapisek Sompote Fund
for Postdoctoral Fellowship, Chulalongkorn University.
\end{acknowledgments}

\bibliographystyle{apsrev4-1}
\bibliography{bib_ref}

\begin{thebibliography}{25}%
\makeatletter
\providecommand \@ifxundefined [1]{%
 \@ifx{#1\undefined}
}%
\providecommand \@ifnum [1]{%
 \ifnum #1\expandafter \@firstoftwo
 \else \expandafter \@secondoftwo
 \fi
}%
\providecommand \@ifx [1]{%
 \ifx #1\expandafter \@firstoftwo
 \else \expandafter \@secondoftwo
 \fi
}%
\providecommand \natexlab [1]{#1}%
\providecommand \enquote  [1]{``#1''}%
\providecommand \bibnamefont  [1]{#1}%
\providecommand \bibfnamefont [1]{#1}%
\providecommand \citenamefont [1]{#1}%
\providecommand \href@noop [0]{\@secondoftwo}%
\providecommand \href [0]{\begingroup \@sanitize@url \@href}%
\providecommand \@href[1]{\@@startlink{#1}\@@href}%
\providecommand \@@href[1]{\endgroup#1\@@endlink}%
\providecommand \@sanitize@url [0]{\catcode `\\12\catcode `\$12\catcode
  `\&12\catcode `\#12\catcode `\^12\catcode `\_12\catcode `\%12\relax}%
\providecommand \@@startlink[1]{}%
\providecommand \@@endlink[0]{}%
\providecommand \url  [0]{\begingroup\@sanitize@url \@url }%
\providecommand \@url [1]{\endgroup\@href {#1}{\urlprefix }}%
\providecommand \urlprefix  [0]{URL }%
\providecommand \Eprint [0]{\href }%
\providecommand \doibase [0]{http://dx.doi.org/}%
\providecommand \selectlanguage [0]{\@gobble}%
\providecommand \bibinfo  [0]{\@secondoftwo}%
\providecommand \bibfield  [0]{\@secondoftwo}%
\providecommand \translation [1]{[#1]}%
\providecommand \BibitemOpen [0]{}%
\providecommand \bibitemStop [0]{}%
\providecommand \bibitemNoStop [0]{.\EOS\space}%
\providecommand \EOS [0]{\spacefactor3000\relax}%
\providecommand \BibitemShut  [1]{\csname bibitem#1\endcsname}%
\let\auto@bib@innerbib\@empty
\bibitem [{\citenamefont {Helbing}(2001)}]{Helbing2001}%
  \BibitemOpen
  \bibfield  {author} {\bibinfo {author} {\bibfnamefont {D.}~\bibnamefont
  {Helbing}},\ }\href {\doibase 10.1103/RevModPhys.73.1067} {\bibfield
  {journal} {\bibinfo  {journal} {Rev. Mod. Phys.}\ }\textbf {\bibinfo {volume}
  {73}},\ \bibinfo {pages} {1067} (\bibinfo {year} {2001})}\BibitemShut
  {NoStop}%
\bibitem [{\citenamefont {Nagatani}(2002)}]{Nagatani2002}%
  \BibitemOpen
  \bibfield  {author} {\bibinfo {author} {\bibfnamefont {T.}~\bibnamefont
  {Nagatani}},\ }\href {http://stacks.iop.org/0034-4885/65/i=9/a=203}
  {\bibfield  {journal} {\bibinfo  {journal} {Reports on Progress in Physics}\
  }\textbf {\bibinfo {volume} {65}},\ \bibinfo {pages} {1331} (\bibinfo {year}
  {2002})}\BibitemShut {NoStop}%
\bibitem [{\citenamefont {Chowdhury}\ \emph {et~al.}(2005)\citenamefont
  {Chowdhury}, \citenamefont {Schadschneider},\ and\ \citenamefont
  {Nishinari}}]{Chowdhury2005}%
  \BibitemOpen
  \bibfield  {author} {\bibinfo {author} {\bibfnamefont {D.}~\bibnamefont
  {Chowdhury}}, \bibinfo {author} {\bibfnamefont {A.}~\bibnamefont
  {Schadschneider}}, \ and\ \bibinfo {author} {\bibfnamefont {K.}~\bibnamefont
  {Nishinari}},\ }\href {\doibase https://doi.org/10.1016/j.plrev.2005.09.001}
  {\bibfield  {journal} {\bibinfo  {journal} {Physics of Life Reviews}\
  }\textbf {\bibinfo {volume} {2}},\ \bibinfo {pages} {318 } (\bibinfo {year}
  {2005})}\BibitemShut {NoStop}%
\bibitem [{\citenamefont {Biham}\ \emph {et~al.}(1992)\citenamefont {Biham},
  \citenamefont {Middleton},\ and\ \citenamefont {Levine}}]{Biham1992}%
  \BibitemOpen
  \bibfield  {author} {\bibinfo {author} {\bibfnamefont {O.}~\bibnamefont
  {Biham}}, \bibinfo {author} {\bibfnamefont {A.~A.}\ \bibnamefont
  {Middleton}}, \ and\ \bibinfo {author} {\bibfnamefont {D.}~\bibnamefont
  {Levine}},\ }\href {\doibase 10.1103/PhysRevA.46.R6124} {\bibfield  {journal}
  {\bibinfo  {journal} {Phys. Rev. A}\ }\textbf {\bibinfo {volume} {46}},\
  \bibinfo {pages} {R6124} (\bibinfo {year} {1992})}\BibitemShut {NoStop}%
\bibitem [{\citenamefont {Tajima}\ and\ \citenamefont
  {Nagatani}(2001)}]{Tajima2001}%
  \BibitemOpen
  \bibfield  {author} {\bibinfo {author} {\bibfnamefont {Y.}~\bibnamefont
  {Tajima}}\ and\ \bibinfo {author} {\bibfnamefont {T.}~\bibnamefont
  {Nagatani}},\ }\href {\doibase https://doi.org/10.1016/S0378-4371(00)00630-0}
  {\bibfield  {journal} {\bibinfo  {journal} {Physica A: Statistical Mechanics
  and its Applications}\ }\textbf {\bibinfo {volume} {292}},\ \bibinfo {pages}
  {545 } (\bibinfo {year} {2001})}\BibitemShut {NoStop}%
\bibitem [{\citenamefont {Newell}(1961)}]{Newell1961}%
  \BibitemOpen
  \bibfield  {author} {\bibinfo {author} {\bibfnamefont {G.~F.}\ \bibnamefont
  {Newell}},\ }\href {\doibase 10.1287/opre.9.2.209} {\bibfield  {journal}
  {\bibinfo  {journal} {Operations Research}\ }\textbf {\bibinfo {volume}
  {9}},\ \bibinfo {pages} {209} (\bibinfo {year} {1961})},\ \Eprint
  {http://arxiv.org/abs/https://doi.org/10.1287/opre.9.2.209}
  {https://doi.org/10.1287/opre.9.2.209} \BibitemShut {NoStop}%
\bibitem [{\citenamefont {{Kai Nagel}}\ and\ \citenamefont {{Michael
  Schreckenberg}}(1992)}]{Nagel1992}%
  \BibitemOpen
  \bibfield  {author} {\bibinfo {author} {\bibnamefont {{Kai Nagel}}}\ and\
  \bibinfo {author} {\bibnamefont {{Michael Schreckenberg}}},\ }\href {\doibase
  10.1051/jp1:1992277} {\bibfield  {journal} {\bibinfo  {journal} {J. Phys. I
  France}\ }\textbf {\bibinfo {volume} {2}},\ \bibinfo {pages} {2221} (\bibinfo
  {year} {1992})}\BibitemShut {NoStop}%
\bibitem [{\citenamefont {Derrida}\ \emph {et~al.}(1992)\citenamefont
  {Derrida}, \citenamefont {Domany},\ and\ \citenamefont
  {Mukamel}}]{Derrida1992}%
  \BibitemOpen
  \bibfield  {author} {\bibinfo {author} {\bibfnamefont {B.}~\bibnamefont
  {Derrida}}, \bibinfo {author} {\bibfnamefont {E.}~\bibnamefont {Domany}}, \
  and\ \bibinfo {author} {\bibfnamefont {D.}~\bibnamefont {Mukamel}},\ }\href
  {\doibase 10.1007/BF01050430} {\bibfield  {journal} {\bibinfo  {journal}
  {Journal of Statistical Physics}\ }\textbf {\bibinfo {volume} {69}},\
  \bibinfo {pages} {667} (\bibinfo {year} {1992})}\BibitemShut {NoStop}%
\bibitem [{\citenamefont {Prigogine}\ and\ \citenamefont
  {Andrews}(1960)}]{Prigogine1960}%
  \BibitemOpen
  \bibfield  {author} {\bibinfo {author} {\bibfnamefont {I.}~\bibnamefont
  {Prigogine}}\ and\ \bibinfo {author} {\bibfnamefont {F.~C.}\ \bibnamefont
  {Andrews}},\ }\href {\doibase 10.1287/opre.8.6.789} {\bibfield  {journal}
  {\bibinfo  {journal} {Operations Research}\ }\textbf {\bibinfo {volume}
  {8}},\ \bibinfo {pages} {789} (\bibinfo {year} {1960})},\ \Eprint
  {http://arxiv.org/abs/https://doi.org/10.1287/opre.8.6.789}
  {https://doi.org/10.1287/opre.8.6.789} \BibitemShut {NoStop}%
\bibitem [{\citenamefont {Lighthill}\ and\ \citenamefont
  {Whitham}(1955)}]{Lighthill1955}%
  \BibitemOpen
  \bibfield  {author} {\bibinfo {author} {\bibfnamefont {M.~J.}\ \bibnamefont
  {Lighthill}}\ and\ \bibinfo {author} {\bibfnamefont {G.~B.}\ \bibnamefont
  {Whitham}},\ }\href {\doibase 10.1098/rspa.1955.0089} {\bibfield  {journal}
  {\bibinfo  {journal} {Proc. R. Soc. Lond. A}\ }\textbf {\bibinfo {volume}
  {229}},\ \bibinfo {pages} {317} (\bibinfo {year} {1955})}\BibitemShut
  {NoStop}%
\bibitem [{\citenamefont {Flynn}\ \emph {et~al.}(2009)\citenamefont {Flynn},
  \citenamefont {Kasimov}, \citenamefont {Nave}, \citenamefont {Rosales},\ and\
  \citenamefont {Seibold}}]{Flynn2009}%
  \BibitemOpen
  \bibfield  {author} {\bibinfo {author} {\bibfnamefont {M.~R.}\ \bibnamefont
  {Flynn}}, \bibinfo {author} {\bibfnamefont {A.~R.}\ \bibnamefont {Kasimov}},
  \bibinfo {author} {\bibfnamefont {J.-C.}\ \bibnamefont {Nave}}, \bibinfo
  {author} {\bibfnamefont {R.~R.}\ \bibnamefont {Rosales}}, \ and\ \bibinfo
  {author} {\bibfnamefont {B.}~\bibnamefont {Seibold}},\ }\href {\doibase
  10.1103/PhysRevE.79.056113} {\bibfield  {journal} {\bibinfo  {journal} {Phys.
  Rev. E}\ }\textbf {\bibinfo {volume} {79}},\ \bibinfo {pages} {056113}
  (\bibinfo {year} {2009})}\BibitemShut {NoStop}%
\bibitem [{\citenamefont {Lee}\ \emph {et~al.}(2001)\citenamefont {Lee},
  \citenamefont {Lee},\ and\ \citenamefont {Kim}}]{Lee2001}%
  \BibitemOpen
  \bibfield  {author} {\bibinfo {author} {\bibfnamefont {H.~K.}\ \bibnamefont
  {Lee}}, \bibinfo {author} {\bibfnamefont {H.-W.}\ \bibnamefont {Lee}}, \ and\
  \bibinfo {author} {\bibfnamefont {D.}~\bibnamefont {Kim}},\ }\href {\doibase
  10.1103/PhysRevE.64.056126} {\bibfield  {journal} {\bibinfo  {journal} {Phys.
  Rev. E}\ }\textbf {\bibinfo {volume} {64}},\ \bibinfo {pages} {056126}
  (\bibinfo {year} {2001})}\BibitemShut {NoStop}%
\bibitem [{\citenamefont {John}\ \emph {et~al.}(2009)\citenamefont {John},
  \citenamefont {Schadschneider}, \citenamefont {Chowdhury},\ and\
  \citenamefont {Nishinari}}]{John2009}%
  \BibitemOpen
  \bibfield  {author} {\bibinfo {author} {\bibfnamefont {A.}~\bibnamefont
  {John}}, \bibinfo {author} {\bibfnamefont {A.}~\bibnamefont
  {Schadschneider}}, \bibinfo {author} {\bibfnamefont {D.}~\bibnamefont
  {Chowdhury}}, \ and\ \bibinfo {author} {\bibfnamefont {K.}~\bibnamefont
  {Nishinari}},\ }\href {\doibase 10.1103/PhysRevLett.102.108001} {\bibfield
  {journal} {\bibinfo  {journal} {Phys. Rev. Lett.}\ }\textbf {\bibinfo
  {volume} {102}},\ \bibinfo {pages} {108001} (\bibinfo {year}
  {2009})}\BibitemShut {NoStop}%
\bibitem [{\citenamefont {Helbing}\ \emph {et~al.}(2000)\citenamefont
  {Helbing}, \citenamefont {Farkas},\ and\ \citenamefont
  {Vicsek}}]{Helbing2000}%
  \BibitemOpen
  \bibfield  {author} {\bibinfo {author} {\bibfnamefont {D.}~\bibnamefont
  {Helbing}}, \bibinfo {author} {\bibfnamefont {I.}~\bibnamefont {Farkas}}, \
  and\ \bibinfo {author} {\bibfnamefont {T.}~\bibnamefont {Vicsek}},\ }\href
  {\doibase 10.1038/35035023} {\bibfield  {journal} {\bibinfo  {journal}
  {Nature}\ }\textbf {\bibinfo {volume} {407}},\ \bibinfo {pages} {487}
  (\bibinfo {year} {2000})}\BibitemShut {NoStop}%
\bibitem [{\citenamefont {Klumpp}\ and\ \citenamefont
  {Lipowsky}(2003)}]{Klumpp2003}%
  \BibitemOpen
  \bibfield  {author} {\bibinfo {author} {\bibfnamefont {S.}~\bibnamefont
  {Klumpp}}\ and\ \bibinfo {author} {\bibfnamefont {R.}~\bibnamefont
  {Lipowsky}},\ }\href {\doibase 10.1023/A:1025778922620} {\bibfield  {journal}
  {\bibinfo  {journal} {Journal of Statistical Physics}\ }\textbf {\bibinfo
  {volume} {113}},\ \bibinfo {pages} {233} (\bibinfo {year}
  {2003})}\BibitemShut {NoStop}%
\bibitem [{\citenamefont {Arita}\ \emph {et~al.}(2017)\citenamefont {Arita},
  \citenamefont {Foulaadvand},\ and\ \citenamefont {Santen}}]{Arita2017}%
  \BibitemOpen
  \bibfield  {author} {\bibinfo {author} {\bibfnamefont {C.}~\bibnamefont
  {Arita}}, \bibinfo {author} {\bibfnamefont {M.~E.}\ \bibnamefont
  {Foulaadvand}}, \ and\ \bibinfo {author} {\bibfnamefont {L.}~\bibnamefont
  {Santen}},\ }\href {\doibase 10.1103/PhysRevE.95.032108} {\bibfield
  {journal} {\bibinfo  {journal} {Phys. Rev. E}\ }\textbf {\bibinfo {volume}
  {95}},\ \bibinfo {pages} {032108} (\bibinfo {year} {2017})}\BibitemShut
  {NoStop}%
\bibitem [{\citenamefont {Woelki}(2013)}]{Woelki2013}%
  \BibitemOpen
  \bibfield  {author} {\bibinfo {author} {\bibfnamefont {M.}~\bibnamefont
  {Woelki}},\ }\href {\doibase 10.1103/PhysRevE.87.062818} {\bibfield
  {journal} {\bibinfo  {journal} {Phys. Rev. E}\ }\textbf {\bibinfo {volume}
  {87}},\ \bibinfo {pages} {062818} (\bibinfo {year} {2013})}\BibitemShut
  {NoStop}%
\bibitem [{\citenamefont {Konishi}\ \emph {et~al.}(2000)\citenamefont
  {Konishi}, \citenamefont {Kokame},\ and\ \citenamefont
  {Hirata}}]{Konishi2000}%
  \BibitemOpen
  \bibfield  {author} {\bibinfo {author} {\bibfnamefont {K.}~\bibnamefont
  {Konishi}}, \bibinfo {author} {\bibfnamefont {H.}~\bibnamefont {Kokame}}, \
  and\ \bibinfo {author} {\bibfnamefont {K.}~\bibnamefont {Hirata}},\ }in\
  \href {\doibase 10.1109/COC.2000.873957} {\emph {\bibinfo {booktitle} {2000
  2nd International Conference. Control of Oscillations and Chaos. Proceedings
  (Cat. No.00TH8521)}}},\ Vol.~\bibinfo {volume} {2}\ (\bibinfo {year} {2000})\
  pp.\ \bibinfo {pages} {221--224}\BibitemShut {NoStop}%
\bibitem [{\citenamefont {Yokoya}(2004)}]{Yokoya2004}%
  \BibitemOpen
  \bibfield  {author} {\bibinfo {author} {\bibfnamefont {Y.}~\bibnamefont
  {Yokoya}},\ }\href {\doibase 10.1103/PhysRevE.69.016121} {\bibfield
  {journal} {\bibinfo  {journal} {Phys. Rev. E}\ }\textbf {\bibinfo {volume}
  {69}},\ \bibinfo {pages} {016121} (\bibinfo {year} {2004})}\BibitemShut
  {NoStop}%
\bibitem [{\citenamefont {Stern}\ \emph {et~al.}(2018)\citenamefont {Stern},
  \citenamefont {Cui}, \citenamefont {Monache}, \citenamefont {Bhadani},
  \citenamefont {Bunting}, \citenamefont {Churchill}, \citenamefont {Hamilton},
  \citenamefont {Haulcy}, \citenamefont {Pohlmann}, \citenamefont {Wu},
  \citenamefont {Piccoli}, \citenamefont {Seibold}, \citenamefont {Sprinkle},\
  and\ \citenamefont {Work}}]{Stern2018}%
  \BibitemOpen
  \bibfield  {author} {\bibinfo {author} {\bibfnamefont {R.~E.}\ \bibnamefont
  {Stern}}, \bibinfo {author} {\bibfnamefont {S.}~\bibnamefont {Cui}}, \bibinfo
  {author} {\bibfnamefont {M.~L.~D.}\ \bibnamefont {Monache}}, \bibinfo
  {author} {\bibfnamefont {R.}~\bibnamefont {Bhadani}}, \bibinfo {author}
  {\bibfnamefont {M.}~\bibnamefont {Bunting}}, \bibinfo {author} {\bibfnamefont
  {M.}~\bibnamefont {Churchill}}, \bibinfo {author} {\bibfnamefont
  {N.}~\bibnamefont {Hamilton}}, \bibinfo {author} {\bibfnamefont
  {R.}~\bibnamefont {Haulcy}}, \bibinfo {author} {\bibfnamefont
  {H.}~\bibnamefont {Pohlmann}}, \bibinfo {author} {\bibfnamefont
  {F.}~\bibnamefont {Wu}}, \bibinfo {author} {\bibfnamefont {B.}~\bibnamefont
  {Piccoli}}, \bibinfo {author} {\bibfnamefont {B.}~\bibnamefont {Seibold}},
  \bibinfo {author} {\bibfnamefont {J.}~\bibnamefont {Sprinkle}}, \ and\
  \bibinfo {author} {\bibfnamefont {D.~B.}\ \bibnamefont {Work}},\ }\href
  {\doibase https://doi.org/10.1016/j.trc.2018.02.005} {\bibfield  {journal}
  {\bibinfo  {journal} {Transportation Research Part C: Emerging Technologies}\
  }\textbf {\bibinfo {volume} {89}},\ \bibinfo {pages} {205} (\bibinfo {year}
  {2018})}\BibitemShut {NoStop}%
\bibitem [{\citenamefont {Bando}\ \emph
  {et~al.}(1995{\natexlab{a}})\citenamefont {Bando}, \citenamefont {Hasebe},
  \citenamefont {Nakanishi}, \citenamefont {Nakayama}, \citenamefont
  {Shibata},\ and\ \citenamefont {Sugiyama}}]{Bando1995}%
  \BibitemOpen
  \bibfield  {author} {\bibinfo {author} {\bibfnamefont {M.}~\bibnamefont
  {Bando}}, \bibinfo {author} {\bibfnamefont {K.}~\bibnamefont {Hasebe}},
  \bibinfo {author} {\bibfnamefont {K.}~\bibnamefont {Nakanishi}}, \bibinfo
  {author} {\bibfnamefont {A.}~\bibnamefont {Nakayama}}, \bibinfo {author}
  {\bibfnamefont {A.}~\bibnamefont {Shibata}}, \ and\ \bibinfo {author}
  {\bibfnamefont {Y.}~\bibnamefont {Sugiyama}},\ }\href {\doibase
  10.1051/jp1:1995206} {\bibfield  {journal} {\bibinfo  {journal} {{Journal de
  Physique I}}\ }\textbf {\bibinfo {volume} {5}},\ \bibinfo {pages} {1389}
  (\bibinfo {year} {1995}{\natexlab{a}})}\BibitemShut {NoStop}%
\bibitem [{\citenamefont {Bando}\ \emph
  {et~al.}(1995{\natexlab{b}})\citenamefont {Bando}, \citenamefont {Hasebe},
  \citenamefont {Nakayama}, \citenamefont {Shibata},\ and\ \citenamefont
  {Sugiyama}}]{Bando1995PRE}%
  \BibitemOpen
  \bibfield  {author} {\bibinfo {author} {\bibfnamefont {M.}~\bibnamefont
  {Bando}}, \bibinfo {author} {\bibfnamefont {K.}~\bibnamefont {Hasebe}},
  \bibinfo {author} {\bibfnamefont {A.}~\bibnamefont {Nakayama}}, \bibinfo
  {author} {\bibfnamefont {A.}~\bibnamefont {Shibata}}, \ and\ \bibinfo
  {author} {\bibfnamefont {Y.}~\bibnamefont {Sugiyama}},\ }\href {\doibase
  10.1103/PhysRevE.51.1035} {\bibfield  {journal} {\bibinfo  {journal} {Phys.
  Rev. E}\ }\textbf {\bibinfo {volume} {51}},\ \bibinfo {pages} {1035}
  (\bibinfo {year} {1995}{\natexlab{b}})}\BibitemShut {NoStop}%
\bibitem [{\citenamefont {Tadaki}\ \emph {et~al.}(1998)\citenamefont {Tadaki},
  \citenamefont {Kikuchi}, \citenamefont {Sugiyama},\ and\ \citenamefont
  {Yukawa}}]{Tadaki1998}%
  \BibitemOpen
  \bibfield  {author} {\bibinfo {author} {\bibfnamefont {S.}~\bibnamefont
  {Tadaki}}, \bibinfo {author} {\bibfnamefont {M.}~\bibnamefont {Kikuchi}},
  \bibinfo {author} {\bibfnamefont {Y.}~\bibnamefont {Sugiyama}}, \ and\
  \bibinfo {author} {\bibfnamefont {S.}~\bibnamefont {Yukawa}},\ }\href
  {\doibase 10.1143/JPSJ.67.2270} {\bibfield  {journal} {\bibinfo  {journal}
  {Journal of the Physical Society of Japan}\ }\textbf {\bibinfo {volume}
  {67}},\ \bibinfo {pages} {2270} (\bibinfo {year} {1998})},\ \Eprint
  {http://arxiv.org/abs/https://doi.org/10.1143/JPSJ.67.2270}
  {https://doi.org/10.1143/JPSJ.67.2270} \BibitemShut {NoStop}%
\bibitem [{Note1()}]{Note1}%
  \BibitemOpen
  \bibinfo {note} {We note that, in real driving experience, a short response
  time is required for safety and comfortable driving.}\BibitemShut {Stop}%
\bibitem [{\citenamefont {Sugiyama}\ \emph {et~al.}(2008)\citenamefont
  {Sugiyama}, \citenamefont {Fukui}, \citenamefont {Kikuchi}, \citenamefont
  {Hasebe}, \citenamefont {Nakayama}, \citenamefont {Nishinari}, \citenamefont
  {ichi Tadaki},\ and\ \citenamefont {Yukawa}}]{Sugiyama2008}%
  \BibitemOpen
  \bibfield  {author} {\bibinfo {author} {\bibfnamefont {Y.}~\bibnamefont
  {Sugiyama}}, \bibinfo {author} {\bibfnamefont {M.}~\bibnamefont {Fukui}},
  \bibinfo {author} {\bibfnamefont {M.}~\bibnamefont {Kikuchi}}, \bibinfo
  {author} {\bibfnamefont {K.}~\bibnamefont {Hasebe}}, \bibinfo {author}
  {\bibfnamefont {A.}~\bibnamefont {Nakayama}}, \bibinfo {author}
  {\bibfnamefont {K.}~\bibnamefont {Nishinari}}, \bibinfo {author}
  {\bibfnamefont {S.}~\bibnamefont {ichi Tadaki}}, \ and\ \bibinfo {author}
  {\bibfnamefont {S.}~\bibnamefont {Yukawa}},\ }\href
  {http://stacks.iop.org/1367-2630/10/i=3/a=033001} {\bibfield  {journal}
  {\bibinfo  {journal} {New Journal of Physics}\ }\textbf {\bibinfo {volume}
  {10}},\ \bibinfo {pages} {033001} (\bibinfo {year} {2008})}\BibitemShut
  {NoStop}%
\end{thebibliography}%

\end{document}